\pdfoutput=1
\documentclass[aps, reprint, superscriptaddress, nofootinbib, showpacs, floatfix]{revtex4-1}
\usepackage{amsmath, latexsym, amssymb, hyperref, graphicx, color, multirow}
\usepackage[T1]{fontenc}
\definecolor{nicered}{rgb}{.7,.1,.1}
\definecolor{nicegreen}{rgb}{.1,.5,.1}
\definecolor{darkblue}{rgb}{0,0,.5}
\hypersetup{colorlinks, citecolor=nicegreen,linkcolor=nicered, urlcolor=darkblue}

\begin{document}

\title{Perturbativity and mass scales of Left-Right Higgs bosons}

\author{Alessio Maiezza}
\email{amaiezza@ific.uv.es}
\affiliation{IFIC, Universitat de Val\`encia-CSIC, Apt. Correus 22085, E-46071 Val\`encia, Spain}

\author{Miha Nemev\v{s}ek}
\email{miha.nemevsek@ijs.si}
\affiliation{Jo\v{z}ef Stefan Institute, Ljubljana, Slovenia}

\author{Fabrizio Nesti}
\email{fabrizio.nesti@irb.hr}
\affiliation{Ru\dj er Bo\v{s}kovi\'c Institute, Bijeni\v{c}ka cesta 54, 10000, Zagreb, Croatia}

\date{\today}

\begin{abstract} \noindent The scalar sector of the minimal Left-Right model at TeV scale is
  revisited in light of the large quartic coupling needed for a heavy flavor-changing scalar. The
  stability and perturbativity of the effective potential is discussed and merged with constraints
  from low-energy processes. Thus the perturbative level of the Left-Right scale is sharpened. Lower
  limits on the triplet scalars are also derived: the left-handed triplet is bounded by oblique
  parameters, while the doubly-charged right-handed component is limited by the $h \to \gamma
  \gamma, Z \gamma$ decays. Current constraints disfavor their detection as long as $W_R$ is within
  the reach of the LHC.
\end{abstract}

\pacs{12.60.Cn, 12.60.Fr, 12.15.Lk\vspace*{0em}}

\maketitle

%
%
\section{Introduction}

\noindent
The standard model (SM) describes all the known particle interactions and their masses, except for the neutrino that is massless within the model, in contrast with the evidence of neutrino oscillations. An appealing solution is provided by the Left-Right symmetric model (LRSM)~\cite{Pati:1974yy, Senjanovic:1975rk}, proposed to explain the most evident misfeature of the SM, that is the glaringly asymmetric chiral structure of the weak interactions. Incidentally, it has arisen as a complete theory for the origin of neutrino masses~\cite{MohSen}.

Left-Right theories possess a rich phenomenology, naturally embed the seesaw mechanism~\cite{Minkowski, MohSen} and, with the right-handed (RH) scale in the TeV region, provide a potentially dominant contribution to neutrino-less double beta decay ($0\nu 2 \beta$)~\cite{Mohapatra:1980yp, Tello:2010am, Nemevsek:2011aa}. This contribution, due to the RH neutrino ($N$), could even be favored in the light of future cosmological bounds on the neutrino masses.  Keung and Senjanovi\'c (KS)~\cite{Keung:1983uu} proposed a low energy equivalent, where the heavy Majorana neutrino production could reveal LNV at colliders~\cite{GSreview}. Recently, a complementary process~\cite{Gunion:1986im} was analyzed at the LHC~\cite{Maiezza:2015lza}, looking for LNV in the SM-like Higgs decay.

The LRSM is based on the gauge group $\mathcal G_{LR} = SU(2)_L \times SU(2)_R \times U(1)_{B-L}$ plus either a generalized parity $\mathcal{P}$ or charge conjugation $\mathcal{C}$~\cite{Maiezza:2010ic}. The model is predictive in a number of ways: the Dirac neutrino coupling is univocally determined from the light and heavy neutrino mass~\cite{Nemevsek:2012iq}, with direct consequences for LHC and for the electric dipole moment of the electron~\cite{Nemevsek:2012iq}. Moreover, in case of $\mathcal P$, the strong CP phase~\cite{Maiezza:2014ala} and the quark flavor mixing in the RH sector are computable from the standard Cabibbo-Kobayashi-Maskawa matrix~\cite{Senjanovic:2014pva}.

However, the TeV scale LRSM is non-trivially bounded by the low energy constraints, in particular $K^0-\overline{K}^0$ and $B_{d,s}^0-\overline{B}^0_{d,s}$ oscillations~~\cite{Beall:1981ze, Ecker:1985vv, Zhang:2007da, Maiezza:2010ic, Bertolini:2014sua}. These, together with CP-violating processes, such as $\varepsilon', \varepsilon$~\cite{Bertolini:2012pu} and the electric dipole moment of the neutron~\cite{Maiezza:2014ala}, set a lower limit on the mass of the RH gauge boson $M_{W_{R}} > 2.9 \text{ TeV}$. In the case of $\mathcal{P}$, the latter can set a substantial bound $M_{W_{R}} > 20 \text{ TeV}$ in case the strong CP problem is taken into account within the model~\cite{Maiezza:2014ala}.

Most of the above limits are in fact dominated by the tree-level exchange of flavor changing (FC) scalars~\cite{Senjanovic:1979cta}, which have to be heavy. Their heavy masses require potentially large quartics that may become non-perturbative. This might be considered as a weak point of the minimal LRSM (see e.g.~\cite{Guadagnoli:2010sd}), apparently spoiling the appeal of a predictive theory of neutrino masses.

The scalar potential of the LRSM with spontaneous parity breaking~\cite{Senjanovic:1975rk} has been the object of study from the original works to subsequent~\cite{Gunion:1989in, Kiers:2002cz, Deshpande:1990ip, Duka:1999uc, Khasanov:2001tu} and recent studies~\cite{Dekens:2014ina, Bambhaniya:2015wna, Dev:2016dja}. Nevertheless, a perturbativity analysis of the scalar potential was missing in literature until now. We provide this missing piece with a loop analysis and a renormalization of the Higgs sector. This leads to an effective potential and to a determination of the parameter space favored by perturbativity.  In turn, it leads to a refined limit on the masses of $W_R$ and the FC scalars. The entire analysis below holds for both the cases of $\mathcal{P}$ and $\mathcal{C}$.

Furthermore, we assess the domain of perturbative regime for the mixing between the RH Higgs boson $\Delta_{R}^{0}$ and the SM-like one, within the phenomenologically interesting region for collider searches~\cite{Maiezza:2015lza}.

Finally, the fairly large FC-scalar quartic also has an impact on oblique parameters and on the SM Higgs diphoton rate. These in turn set a lower bound on the left-handed triplet multiplet and the doubly charged component of the RH one. Therefore, an observation of $W_R$ at the LHC would practically exclude the production of the entire left-handed triplet multiplet in the minimal LRSM. It would also allow for marginal space to observe the RH doubly charged component. On the other hand, the neutral RH Higgs remains fairly unconstrained and should be searched for at the LHC.

%
%
\section{Scalar Potential(s) at tree-level} \label{secScalPot}

\noindent
The scalar content of the LRSM consists of a bi-doublet and two triplets
\begin{equation}
\phi = \begin{pmatrix}
  \phi_1^0 & \phi_2^+ \\ \phi_1^- & \phi_2^0
\end{pmatrix},
\quad
\Delta_{L,R} = \begin{pmatrix}
 \frac{\Delta^{+}}{\sqrt{2}} & \Delta^{++} \\ \Delta^0 & -\frac{\Delta^{+}}{\sqrt{2}}
\end{pmatrix}_{L, R}.
\end{equation}
With the quantum number assignment $\phi \in \left(2_L,2_R,0 \right)$, $\Delta_{L(R)} \in
\left(3_L(1_L),1_R(3_R),2_{B-L}\right)$ under $\mathcal G_{LR}$, the most general
potential ($\mathcal V$) is constructed~\cite{Senjanovic:1975rk}. Since the basic feature of the
LRSM is restoration of parity, an additional restriction is imposed on $\mathcal V$ for the case of
$\mathcal P$ and $\mathcal C$. See Eqs.~\eqref{eqVP} in the Appendix for the most general
potentials.

Both gauge symmetry and LR parity are broken spontaneously (SSB) via the vev of the scalar fields
\begin{equation} \begin{split}
\langle \phi \rangle &= \begin{pmatrix}
  v_1 & 0 \\ 0 & v_2 \, e^{i \alpha}
\end{pmatrix}, \,
\\
\langle \Delta _{L} \rangle &= \begin{pmatrix}
 0 & 0 \\ v _{L}e^{i \theta_L} & 0
\end{pmatrix}, \quad
\langle \Delta _{R} \rangle = \begin{pmatrix}
 0 & 0 \\ v _{R} & 0
\end{pmatrix}, \end{split}
\end{equation}
where $v^2 \equiv v_1^2 + v_2^2 = 174 \text{ GeV}$ and $x \equiv \tan \beta = v_2 / v_1 < 1$. Due to
phenomenological constraints, the scales set by the vevs are fairly hierarchical $v_L \ll v \ll
v_R$. For future convenience, it is useful to introduce the small parameter $\varepsilon = v/v_R$.

The symmetry breaking follows the pattern $\mathcal G_{LR} \xrightarrow{v_R} SU(2)_L\times U(1)_{Y}
\xrightarrow{v_{1,2}} U(1)_{e.m.}$, and various masses are generated spontaneously. In particular,
the gauge boson masses are
\begin{equation}\label{eqMWR}
  M_{W_R} \simeq g \, v_R\,, \qquad M_{W} \simeq  \frac{g \, v}{\sqrt{2}}\,.
\end{equation}
In the following sections, we briefly review the minimization of the potential and the generation of the scalar mass spectrum of the LRSM.

%
\subsection{Minimization of $\mathcal V_{\mathcal P, \mathcal C}$}
\noindent
The minimization equations can be written as
\begin{equation} \label{eqDV}
  \partial_{v_i} \mathcal{V} = 0\, ,
\end{equation}
for all $v_i \in \{v_1, v_2, v_R, v_L, \theta_L, \alpha \}$. In what follows, we stick to the
$\mathcal{P}$ potential. The case of $\mathcal{C}$ follows straightforwardly and the results are
summarized in the Appendix~\ref{secAppPot}.

The first three conditions in Eq.~\eqref{eqDV} provide $\mu_i^2$ as a function of the vevs and the quartic couplings, shown in~\eqref{eqmu12P}-\eqref{eqmu32P}. The derivative on $\alpha$ provides a relation~\eqref{eqRelD2} between the CP phases~\cite{Kiers:2002cz}, which for small $\varepsilon$ and $x$ reads
\begin{equation} \label{eqRelD2Exp}
  2 \alpha_2 \sin \delta_2 \simeq \alpha_3 x \sin \alpha\,.
\end{equation}
Finally, derivation over $v_L$ gives the well-know seesaw relation
\begin{equation} \label{eqVevssaw}
  \begin{split}
  v_L =&\, \frac{\varepsilon^2 v_R}{\left(1+x^2\right) \left(2 \rho_1 - \rho_3\right)} \biggl[
    \beta_1 x \cos (\alpha - \theta_L)
    \\
    &+ \beta_2 \cos (\theta_L) + \beta_3 x^2 \cos (2 \alpha - \theta_L) \biggr] \,.
  \end{split}
\end{equation}

The phenomenology of neutrinos in the LRSM prevents $v_L$ from taking a too large value
\begin{equation} \label{eqvllimit}
  v_L < \frac{m_\nu}{m_N} v_R \simeq 10^{-5} \text{ GeV} \left(\frac{100 \text{ MeV}}{m_N}\right) \left(\frac{v_R}{10 \text{ TeV}} \right),
\end{equation}
because $m_N \gtrsim 100 \text{ MeV}$ due to constraints from supernovae~\cite{Barbieri:1988av} and
Big Bang nucleosynthesis~\cite{Nemevsek:2011aa}, apart from the possible keV DM
candidate~\cite{Nemevsek:2012cd}.

This requires fairly small $\beta_i \lesssim 10^{-4}$, which are technically natural, both from fermion loops because of small neutrino Dirac masses, and from scalar loops since they are self-proportional. Even so, a stabilizing symmetry can be imposed~ \cite{Gunion:1989in, Deshpande:1990ip, Khasanov:2001tu} to guarantee a small $v_L$.

The remaining minimization condition on the derivative with respect to $\theta_L$ is automatically satisfied when $v_L\rightarrow 0$. In any case, for the purpose of this work $\beta_i$ play no significant role, hence we drop them from here on.

%
\subsection{Masses and physical states}

\noindent
Let us now construct the Hessian of $\mathcal{V}$ and use the minimization solutions in~\eqref{eqmu12P}-\eqref{eqmu32P} and~\eqref{eqVevssaw}. The positivity condition on the potential requires positive eigenvalues of the Hessian, which corresponds to positive squared masses of all the scalars. In this way, $8 \times 8$, $4\times4$ and $2\times 2$ matrices for squared masses of the neutral, singly and doubly charged fields are obtained.

At zero order $(\epsilon, x \to 0)$ these matrices are already diagonal. At higher orders in $\varepsilon$, subleading off-diagonal terms appear and one can solve the eigensystem perturbatively at given order, taking care of the would-be-Goldstone components. $\Delta_L$ decouples in the limit of $v_L \rightarrow 0$.

At first order in $\varepsilon$ and $\mathcal O\left(x^2, \varepsilon\,x \right)$, we get the SM Higgs
\begin{equation}\label{SMHiggs}
 h = \text{Re} \, \phi_1 + x \, \text{Re} \, \phi_2 - \theta \, \text{Re} \, \Delta_R^0\,,
\end{equation}
with the EW mass
\begin{equation}\label{mh}
  m_h^2 = \left(4 \lambda_1 - \frac{\alpha_1^2}{\rho_1}\right) v^2\,,
\end{equation}
and the mixing with the neutral component of $\text{Re} \Delta_R^0$ as\footnote{In phenomenological
  applications, the simplified $2 \times 2$ mass matrix with only $h, \Delta_R^0$ is reliable in the $x \ll 1$ limit, e.g.\
  as in~\cite{Maiezza:2015lza}. In this case, the exact mixing parameter is $\tan \left(2 \theta
  \right ) = \frac{\alpha_1 v v_R}{\rho_1 v_R^2 - \lambda_1 v^2}$, from which the upper limit
  $\left| \theta \right| < \pi/4$ is clear.}
\begin{equation}\label{theta}
  \theta = \varepsilon \frac{\alpha_1}{2 \rho_1}\,.
\end{equation}
All the other scalars have the leading mass terms proportional to $v_R$ with subleading electroweak
corrections; the spectrum is shown in Table~\ref{scalarP}. The same result holds for the case of
$\mathcal{C}$, except for the masses of the FC scalars, quoted in the Appendix~\ref{secAppPot},
where the extra phases present in the potential cause modifications.

\begin{table}
\renewcommand{\arraystretch}{1.8}
\begin{tabular}{| l | l | l |}
\hline\hline
 &  $\text{mass}^2$ in $v_R^2$ units &  states \\ \hline
 \multirow{2}{*}{$H$}
 &
 \multirow{2}{*}{$\alpha_3 + 4 \varepsilon^2 \left( 2 \lambda_2 + \lambda_3 + \frac{4 \alpha_2^2}{\alpha_3-4 \rho_1} \right)$}
 &  $\text{Re} \, \phi_2 - x \, \text{Re} \, \phi_1$
 \\
 & & $ \quad+ \varepsilon \frac{4 \alpha_2}{\alpha_3-4 \rho_1}\text{Re} \, \Delta_R^0$
 \\ \hline
 $H'$ &  $ \alpha_3 + 9 \varepsilon^2 (\lambda_3-2 \lambda_2)$  & $\text{Im} \, \phi_2 + x \, \text{Im} \, \phi_1$
 \\ \hline
 $H^+$ & $  \alpha_3 + \varepsilon^2 \frac{\alpha_3}{2} $   &  $\phi_2^+ + x \, \phi_1^+ + \frac{\varepsilon}{\sqrt{2}} \Delta_R^+$
 \\ \hline \hline
 \multirow{2}{*}{$\Delta_R^0$} &
 \multirow{2}{*}{$4 \rho_1 +  \varepsilon^2 \left(\frac{\alpha_1^2}{\rho_1} - \frac{16 \alpha_2^2}{\alpha_3 - 4 \rho_1} \right)$}
 &
 $\text{Re} \, \Delta_R^0 + \varepsilon \frac{\alpha_1}{2 \rho_1}\text{Re} \, \phi_1$
 \\
 & & $\quad - \varepsilon \frac{4 \alpha_2}{\alpha_3-4 \rho_1}\text{Re} \, \phi_2$
 \\\hline
$\Delta_R^{++}$    & $ 4 \rho_2 + \varepsilon^2  \alpha_3$              &         $  \Delta_R^{++} $
  \\ \hline \hline
 $\Delta_L^0$ &   $\rho_3 - 2 \rho_1$       &   $\text{Re} \, \Delta_L^0 \text{ and Im} \, \Delta_L^0$
  \\ \hline
$\Delta_L^{+}$     & $ \rho_3-2 \rho_1 + \varepsilon^2 \frac{\alpha_3}{2} $             &    $  \Delta_L^{+} $
  \\ \hline
$\Delta_L^{++}$    & $ \rho_3-2 \rho_1 + \varepsilon^2 \alpha_3 $             &    $  \Delta_L^{++} $
  \\ \hline\hline
\end{tabular}
\caption{Masses and states in the LRSM ($\mathcal{P}$).
 \label{scalarP}}
\end{table}

%
%
\subsection{Flavor changing effects}

\noindent
The Yukawa couplings of $H$ and $H'$ to SM fermions lead to a non-diagonal Lagrangian in the flavor space~\cite{Senjanovic:1979cta}.
\footnote{A comment on possible FC effects of the SM-like Higgs boson is in order. The state shown in~\eqref{SMHiggs} leads to a diagonal Yukawa couplings in the Lagrangian, reproducing the SM. Expanding~\eqref{SMHiggs} to higher orders, FC effects occur because of small mixing with $H(H')$ of order $m_h^2/m_{H,H'}^2 \approx 10^{-5}$, however the related phenomenology is beyond current experimental search.}
As a consequence, FC processes such as meson oscillations, occur at tree-level and this sets the masses of  $H(H')$ to be fairly heavy. A well known example is the one of $K^0-\bar{K}^0$ mixing~\cite{Beall:1981ze, Ecker:1985vv, Zhang:2007da, Maiezza:2010ic} where $H, H'$ and $W_R$, through the usual box diagrams, contribute to the process. As a result, the bound on the model manifests as a correlated constraint between $M_{W_{R}}$ and $m_{H,H'}$.

Along this line, it was shown in~\cite{Bertolini:2014sua} that constraints from
$B_{d,s}^0-\bar{B}_{d,s}^0$ are even more stringent than the ones from the Kaon sector, partially
due to loop corrections to the tree-level FC contribution.  From~\eqref{eqMWR} and the leading
masses of $H,H'$ in Table~\ref{scalarP},
\begin{equation}
m_{H,H'} \simeq \sqrt{\alpha_3} \, v_R\,,
\end{equation}
it is clear that a fairly large $\alpha_3$ is required for a low RH scale to be compatible with $B$
mixing.  We incidentally add that a sizeable $x$ reduces $\alpha_3$ for a given $m_H$ in the above
mass relation, but simultaneously increases the couplings to fermions, thus worsening the
perturbativity of $\alpha_3$~\cite{Senjanovic:1979cta, Maiezza:2010ic}.  In addition, it also
increases the mixing of $H$ with $\Delta_R^0$ and further exacerbates the problem.  Therefore we
stick to $x \ll 1$ in what follows, the general case requires a dedicated study~\cite{GoranJuan16}.

The size of $\alpha_3$ can be approximated as a function of $M_{W_R}$ as~\cite{Bertolini:2014sua}
\begin{equation}\label{hugea3}
  \alpha_3\approx \frac{30}{\left( M_{W_{R}}/\text{2.23\,TeV} \right)^2-1}\,.
\end{equation}

This relation may generate a tension in the LRSM for low scale of $M_{W_R}$, where $\alpha_3$
becomes large and potentially non-perturbative. Therefore, an evaluation of quantum corrections to
the classical potential is in order, and we discuss it in the following section.

%
%
\section{Perturbativity and potential at quantum level} \label{secQuantPot}

\noindent
An early discussion of perturbativity was presented already in~\cite{Senjanovic:1979cta}, and a
rough further estimate was made by~\cite{Basecq:1985cr} in the study of meson oscillations. The
authors estimate the perturbative regime as $m_H < 10 \, M_{W_R}$. In order to update and clarify the perturbative
status of the LRSM, we study the effective potential and focus on terms generated
by the large $\alpha_3$.  In subsection \ref{subsecVeff}, we first renormalize the model and then
build the effective potential. We compare the tree-level vertices with the one-loop correction in
subsection~\ref{subsecPertrb}, where perturbativity constraints on the relevant LRSM parameters
emerge.

%
\subsection{The effective potential} \label{subsecVeff}
\paragraph{Renormalization.} In order to construct the effective potential, one has to renormalize the theory with proper counter-terms. To illustrate the point, we first focus on the vertices related to $\Delta_R^0$ and generalize to other ones below. For the present purpose, we demonstrate that only $\delta \mu_3$ and $\delta \rho_1$ need to be introduced
\begin{equation} \label{Vct}
  \mathcal{V}_{ct}=\delta \mu_3 \left[\Delta_R^{{\color{white}\dagger}} \Delta_R^\dagger \right] +
  \delta \rho_1 \left[\Delta_R^{{\color{white}\dagger}} \Delta_R^\dagger \right]^2\,,
\end{equation}
where the trace is implied by the square parenthesis.

The following renormalization conditions are imposed:
\begin{itemize}
  \item the tadpoles of $\Delta_R^0$ vanish,
  \item the mass of $\Delta_R^0$ remains the tree-level one.
\end{itemize}
This procedure ensures the finiteness of one loop contributions to any $n$-leg interaction of the scalars, delineating the effective potential. It also allows one to focus on interaction terms while keeping masses intact.

\begin{figure}
  \centerline{\includegraphics[width=\columnwidth]{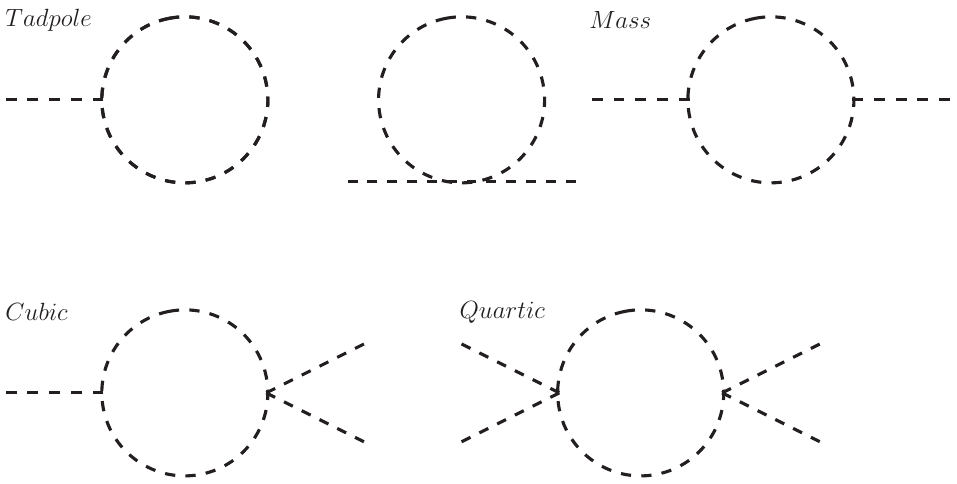}}
  \caption{Divergent diagrams that fix the counter-terms in Eq.~\eqref{Vct}.
  \label{fig:divergent_diagram}\vspace*{-1ex}}
\end{figure}

Let us first focus on the divergent parts of diagrams in Fig.~\ref{fig:divergent_diagram}. It is enough to stick to the diagrams in which heavy scalars with mass dominated by $\alpha_3$ are propagating. The generalization to the loops with the propagating $\Delta_R^0$ is straightforward.

With the above renormalization conditions, the divergent parts of the tadpole and mass diagrams in Fig.~\ref{fig:divergent_diagram} provide the equations
\begin{align} \label{eqTadpole}
  \delta_{\mu_{3}}  - 2 v_R^2 \left(\alpha_3^3 \Delta_\epsilon-\delta_{\rho_{1}} \right) &= 0\,,
  \\ \label{eqMass}
  \delta_{\mu_{3}} -2 v_R^2 \left(\alpha_3^2 (\alpha_3+2) \Delta_\epsilon-3 \delta_{\rho_{1}}  \right) &= 0\,,
\end{align}
where the divergent part is defined as $\Delta_\epsilon = 16\pi^2 \left( 2/\epsilon - \gamma + \log (4\pi) \right)$. These equations fix the counterterms, i.e.\
%
  $\delta_{\rho_{1}} = \alpha_3^2 \Delta_\epsilon$ and
  $\delta_{\mu_{3}}= 2 \alpha_3^2 (\alpha_3-1) v_R^2 \Delta_\epsilon$.
Simulataneously, the divergent part of the three-legs diagram in Fig.~\ref{fig:divergent_diagram},
\begin{equation}
  6 i \sqrt 2 v_R \left(\alpha_3^2 \Delta_\epsilon - \delta_{\rho_1} \right),
\end{equation}
vanishes when $\delta_{\rho_{1}}$ is inserted and leaves a finite result.

\paragraph{The effective potential.} The finiteness holds for any number of external legs in the loop and all the finite parts can be formally summed as~\cite{Peskin}
\begin{equation}\label{formaleffective}
  \mathcal{V}_{eff}=\mathcal{V}+\mathcal{V}_{ct}+\frac{1}{4} \sum_i \frac{m_i^4}{(4 \pi )^2}\left[\log \left(\frac{ m_i^2}{\mu ^2}\right)-\frac{3}{2}\right],
\end{equation}
where $m_i^2$ are the eigenvalues of the Hessian $\frac{\partial^2\mathcal{V}}{\partial\phi_i \partial\phi_j}$ in the unbroken phase. Once the counter-terms are inserted in~\eqref{formaleffective}, a finite expression appears with $\mu-$dependence in the constant part only.

Since we are interested in the effects of a large $\alpha_3$, we restrict the sum in Eq.~\eqref{formaleffective} to $\Delta_R^0$ and $H, H'$ and $H^\pm$. Specifically, the relevant part for $\Delta_R^0$, that is the Higgs of the LRSM, can be expanded for small field values
\begin{align} \label{expanseffective}
 \mathcal V_{eff} =&\, 4 v_R^2 \rho_1{\Delta_R^0}^2
\nonumber
 \\
 &{}+ \left[ 4 \rho_1 \! + \! \frac{2}{(4 \pi)^2} \left( \frac{4}{3}\alpha_3^2 + 18 \rho_1^2 \right) \right] v_R {\Delta_R^0}^3
 \\
 &+ \left[ \rho_1 \! + \! \frac{1}{(4 \pi)^2} \left(\frac{8}{3} \alpha_3^2+27 \rho_1^2 \right)  \right] {\Delta_R^0}^4 + \mathcal O\left(\!{\Delta_R^0}^5\right).
\nonumber
\end{align}

\paragraph{Stability.} The quantum corrections in~\eqref{expanseffective} may be dominated by $\alpha_3$, which also affects the stability of the potential. This assessment requires an expansion of Eq.~\eqref{formaleffective} in the large field limit
\begin{equation} \label{eqEffPotLarge}
  \mathcal{V}_{eff} \approx \frac{1}{32 \pi^2} \alpha_3^2 {\Delta_R^0}^4 \log \left(\frac{{\Delta_R^0}^2}{v_R^2}\right) >0\,,
\end{equation}
showing that stability is enhanced by the large $\alpha_3$.\footnote{In~\cite{Chakrabortty:2013mha} an upper limit on $\alpha_3$ was claimed from the requirement of boundedness of the classical potential. However, we note that the $\alpha_3$ term in the scalar potential is positive definite, therefore no upper bound on $\alpha_3$ can exist. We note that this result was used in a few subsequent works~\cite{UseOfCopositivity}.}

Incidentally, the sizeable $\alpha_3$ contribution in the effective potential relaxes the destabilizing role of RH neutrinos~\cite{Mohapatra:1986pj}, as described below.

\begin{figure}
   \includegraphics[width=\columnwidth]{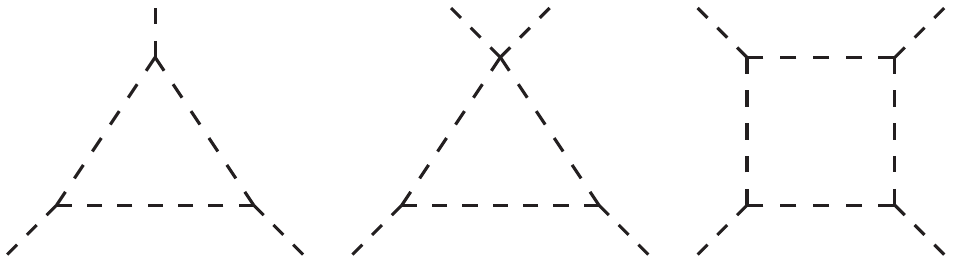}
  \caption{Finite contributions to three and four legs scalar interactions that enter in the measure of perturbativity.
  \label{figFinite}\vspace*{-1ex}}
\end{figure}

%
\subsection{Perturbativity constraints}\label{subsecPertrb}

\noindent
In this section, we assess the level of perturbativity within the LRSM parameter space. Ultimately, one is interested in the mass scales of the theory, therefore we consider those parameters that are responsible for the leading mass contribution in Table~\ref{scalarP}, in particular $\alpha_3$ and also $\rho_{1,2,3}$.

We generalize the renormalization procedure outlined above, include the proper counterterms fixed by diagrams in Fig~\ref{fig:divergent_diagram} and follow the same scheme. Each vertex of $\mathcal V_{eff}$ will receive one loop contributions from many couplings, different from the tree-level one. This leads to correlated bounds, driven by vertices and suppressed by masses. We find that a reliable measure of perturbativity is given by purely self-induced corrections, computed by taking only one large coupling at the time. The dominant ones are four-leg vertices from Figs.~\ref{fig:divergent_diagram} and~\ref{figFinite}, while the three-leg are less stringent. The ratio between the self-generated 1-loop and the corresponding tree-level parameter is then taken as a measure of perturbativity
\begin{align} \label{a3}
  \frac{\alpha_3^{(1)}}{\alpha_3} &= \frac{3 \alpha_3}{8\pi^2}\,,
  \\ \label{r1}
  \frac{\rho_1^{(1)}}{\rho_1} &= \frac{27 \rho_1}{16\pi^2}\,,
  \\
  \frac{\rho_2^{(1)}}{\rho_2} &= \frac{7 \rho_2}{4\pi^2}\,,
  \\ \label{r3}
  \frac{\rho_3^{(1)}}{\rho_3} &= \frac{3 \rho_3}{16\pi^2}\,,
\end{align}
where the superscript $(1)$ denotes the 1-loop value. Note that~\eqref{r1} reproduces the result from the effective potential in~\eqref{expanseffective}, as it should.

In addition to the quartics related to heavy scalars, one can generalize the discussion to the SM-like Higgs $h$. In particular, focusing on the $\lambda_1$ quartic, one gets
\begin{equation} \label{l1}
  \frac{\lambda_1^{(1)}}{\lambda_1} = \frac{27 \lambda_1}{16\pi^2}\,.
\end{equation}

It should be kept in mind that when the actual physical processes (e.g. $\Delta_R^0 \Delta_R^0 H H$ scattering) are considered, the vertices of the interaction will typically be suppressed with respect to those in $\mathcal V_{eff}$. This is due to non-zero masses and momentum flowing in the loop. Both will regulate the loops and make the perturbative expansion more stable as the evaluation of the $\mathcal V_{eff}$.

In addition to perturbativity of the effective potential, one may consider requiring tree level unitarity of the scattering of LRSM scalars. The bounds from the optical theorem imply $\rho_{1,2} < 2 \pi, \, \alpha_{3}, \rho_3 < 8 \pi$. These are close or slightly more stringent than the ones from $100\%$ perturbative level for $\rho_{1,2}, \alpha_3$, while for $\rho_3$ the unitarity bound matches with $\sim 50 \%$ perturbativity.

%
%
\section{Mass scales in TeV LRSM} \label{secImplication}

\noindent
The discussion in the previous section was independent of the LR scale, i.e.\ the perturbativity measure in Eqs.~\eqref{a3}-\eqref{r3} are independent of $M_{W_R}$. Here we apply those results on the mass spectrum of the LRSM scalars.

In subsection~\ref{subsecBMixWR}, we analyze the level of perturbativity for given masses of $H$ and $W_R$, we then consider the mixing $\theta$ of the SM Higgs with $\Delta_R^0$ in~\ref{subsecHiggsMix}, followed by limits on the RH neutrino from the stability of $\mathcal V_{eff}$ in~\ref{subsecN}. Finally, in subsections~\ref{subsecOblqDL} and~\ref{subsecHtogaga}, we deal with implications of $\alpha_3$ on the mass of $\Delta_L$ due to oblique parameters and $\Delta_R^{++}$ via radiative Higgs decays to $\gamma \gamma (Z \gamma)$.

%
\begin{figure}
   \includegraphics[width=\columnwidth]{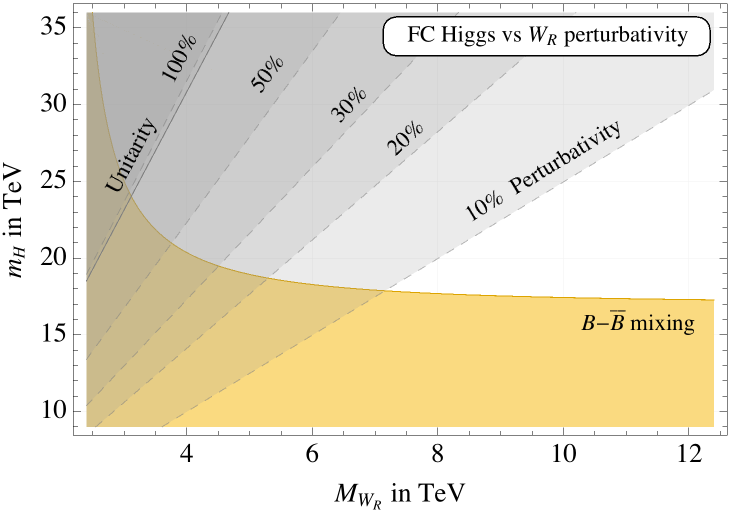}
  \caption{Perturbativity assessment of $\mathcal{V}_{eff}$ (dashed) and tree-level unitarity (solid) of $\alpha_3$, together with the bound on $M_{W_R}$ vs. $m_H$ from $B_{d,s}^0-\overline{B}^0_{d,s}$ (see~\cite{Bertolini:2014sua} for details).}
  \label{mass_scale}
\end{figure}

\subsection{$B$ mixing and $W_R$} \label{subsecBMixWR}

\noindent The estimate of perturbativity and unitarity on $\alpha_3$ from the previous section together with the $B$-oscillations limit in Eq.~\eqref{hugea3}, impose a bound on $M_{W_R}$ for a given perturbative level, as shown on Fig.~\ref{mass_scale}. In~\cite{Bertolini:2014sua} a rough evaluation of such bound for 100\% perturbativity led to $M_{W_R} \gtrsim 3 \text{ TeV}$. As clear from Fig.~\ref{mass_scale}, the improved treatment basically confirms that result with a slight increase.

It is worth noting that the $\alpha_3$ vertices from the scalar potential could directly affect the $B_{d,s}^0-\overline{B}^0_{d,s}$ analysis after the renormalization of the $H$ propagator as in the scheme of~\cite{Basecq:1985cr}, where only gauge interactions were taken into account. The ensuing modification of Eq.~\eqref{hugea3} turns out to be $\sim 2\%$, therefore the $B$ mixing bound remains reliable even with $\alpha_3$ close to the unitarity bound.

%

\subsection{Higgs mixing} \label{subsecHiggsMix}

\noindent
Using the results of the previous section, one can understand the perturbative range of couplings relevant for the $h-\Delta_R^0$ Higgs mixing. In the physically interesting region $m_{\Delta_{R}^{0}} < \mathcal O(\text{TeV})$, its quartic $\rho_1$ is small and thus perturbative. As a result, $\alpha_1$ also turns out to be small, for a given mixing angle, from~\eqref{theta}. The point is then, that the correction to the mass of the Higgs in Eq.~\eqref{mh} becomes sizeable and needs to be canceled by a large $\lambda_1$ in order to preserve the light SM Higgs.

\begin{figure}
  \includegraphics[width=\columnwidth]{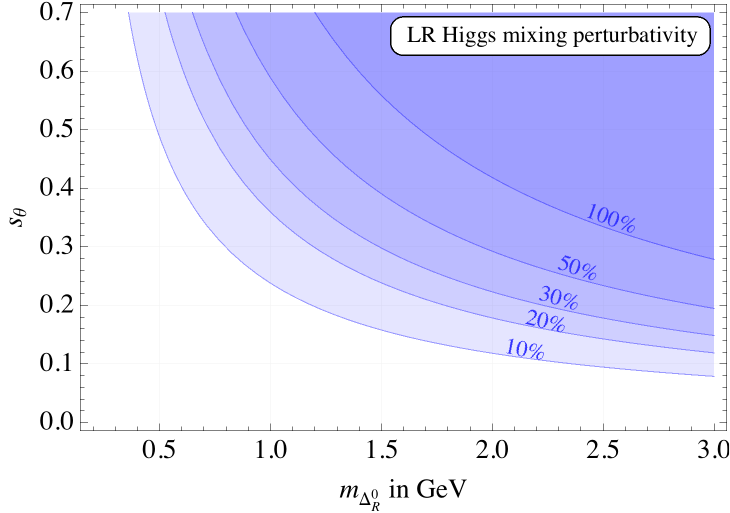}
  \caption{Regions of $\Delta_R^0$--SM Higgs mixing $s_\theta$ favored by $\lambda_1$
    perturbativity.}
  \label{figTheta_Delta}
\end{figure}

To a good approximation $\lambda_1$ can be determined for given $m_{\Delta_R^0}$ and $\theta$ as
\begin{equation}
  4 v^2 \lambda_1 \simeq m_h^2 + \theta^2 m_{\Delta_R^0}^2\,,
\end{equation}
which shows that with significant mixing $\theta$, $\lambda_1$ becomes large for $\Delta_R^0$ in the
TeV range. At this point $\lambda_1$ may clash with Eq.~\eqref{l1}, derived in the small $\alpha_1$
limit, which translates into an upper bound on the mixing for a given $\Delta_R^0$ mass, shown in
Fig.~\ref{figTheta_Delta}. A considerable mixing is allowed for $\Delta_R^0$, whose mass can lie in
the TeV range, still in perturbative regime.

Notice that Higgs mixing will significantly affect electroweak processes and a stringent constraint on $\theta$ vs. the singlet mass was reported in~\cite{singletHiggs}. However, the situation in the LRSM is more involved than the singlet case in~\cite{singletHiggs} due to simultaneous presence of other sources (e.g. $\Delta_L$ below). A complete study would be in order, but lies beyond the scope of this paper.

\subsection{Limit on the RH neutrino mass} \label{subsecN}

\noindent
As mentioned above, the sizeable $\alpha_3$ contribution in the effective potential is positive, and this relaxes the destabilizing role of RH neutrinos. Their impact was considered in~\cite{Mohapatra:1986pj}, where the role of quartics was not considered.
Taking the RH neutrinos into account together with Eq.~\eqref{eqEffPotLarge}, we estimate
\begin{equation}\label{eq:mnbound}
  m_N \lesssim 2^{\frac{1}{4}} \sqrt{\alpha_3} \, v_R \simeq 1.85 \, M_{W_R} \sqrt{\alpha_3}\,,
\end{equation}
where the gauge boson contribution is subleading and can be safely neglected. In fact, the $B$-mixing constraints imply the lower bound on $\alpha_3$ as from Eq.~\eqref{hugea3}. By considering this minimal required value of $\alpha_3$, we find that $m_N$ below 25\,\text{TeV} is allowed by stability, regardless of the LR scale.

\begin{figure}
   \centerline{\includegraphics[width=\columnwidth]{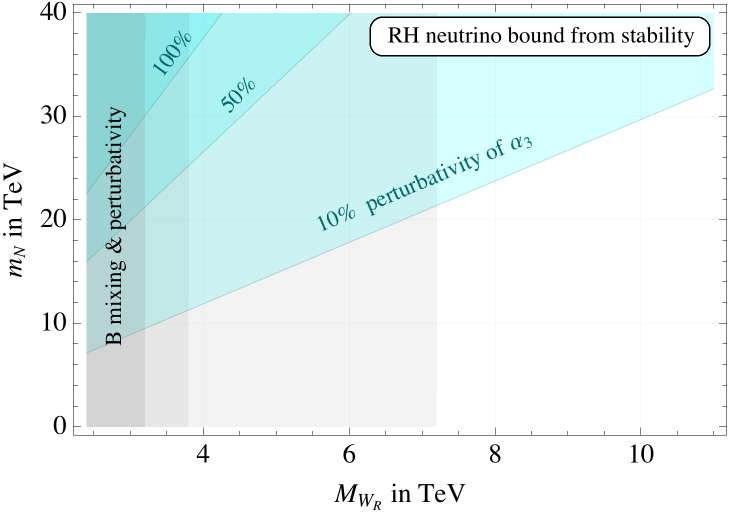}}
   \caption{An upper limit on $m_N$ from the stability of the effective potential. The limit depends
     on $\alpha_3$, which in turn is perturbatively restricted (see text). The gray vertical bands
     correspond to regions disfavored by perturbativity at 10\%, 50\%, 100\% (right to left) as from
     Fig.~\ref{mass_scale}.}
   \label{figmNVeff}
 \end{figure}

The perturbativity upper bound on $\alpha_3$ implies instead an absolute upper bound on $m_N$, depending on the required level of perturbativity. From~\eqref{eq:mnbound} and using the upper values of $\alpha_3$ from Eq.~\eqref{a3}, we have
\begin{equation}
  \frac{m_N}{M_{W_R}} \ \lesssim\  2.3\,,\  5.1\,,\  7.3\,,
\end{equation}
at 10, 50, 100\% perturbativity, as shown on Fig.~\ref{figmNVeff}. It is clear that a quite heavy $m_N$ is allowed, without ruining stability or perturbativity. This bound is further relaxed by the positive contributions of the other quartic couplings to $\mathcal V_{eff}$.

%

%
\subsection{Oblique parameters and $\Delta_L$} \label{subsecOblqDL}

\noindent
Oblique parameters, in particular the $S$ and $T$, impose significant constraints on $SU(2)_L$ multiplets near the EW scale. The isospin violating $T$ parameter is in principle sensitive to $v_L$ and mass splittings within the $\Delta_L$ multiplet~\cite{Gunion:1989in, Melfo:2011nx}. However, due to the smallness of $v_L$ (see Eq.~\eqref{eqvllimit}), only mass splittings are relevant.

In contrast to the stand-alone type II scenario, where the mass splitting is arbitrary, the embedding in the LRSM fixes the size and the sign of the spectrum. From Tab.~\ref{scalarP} the following sum rule is obtained:
\begin{equation} \label{eqDLSumRuleLR}
  \frac{m^2_{\Delta_L^{++}} - m^2_{\Delta_L^+}}{M_W^2} = \frac{m^2_{\Delta_L^+} - m^2_{\Delta_L^0} }{M_W^2}
  = \left( \frac{m_H}{M_{W_R}} \right)^2 > 0\,,
\end{equation}
which is quite robust, with corrections $\mathcal O(v_L/v)^2$.

\medskip

\begin{figure}
   \centerline{\includegraphics[width=\columnwidth]{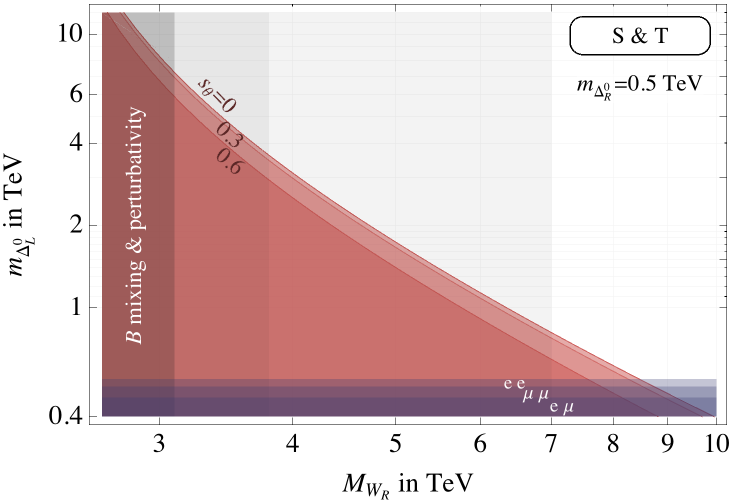}}
   \caption{Lower bound on the mass of the $\Delta_L$ multiplet, dominated by the large $T$
     parameter. The shaded regions are excluded at $2 \sigma$ level for $s_\theta = (0, 0.3, 0.6)$
     (upper to lower). The gray vertical bands correspond to regions disfavored by perturbativity at
     10\%, 50\%, 100\% (right to left) as from Fig.~\ref{mass_scale}. \vspace*{1em}}
  \label{figDeltaLEWPT}
\end{figure}

The splitting in Eq.~\eqref{eqDLSumRuleLR} is set by $\alpha_3$, which in turns depend on $M_{W_R}$, see Eq.~\eqref{hugea3}. Therefore, at low scales, the LRSM requires sizeable mass splittings of $\Delta_L$ components. This has two significant implications:
\begin{enumerate}
  \item The splitting of $\Delta_L$ components induces a large $T$ parameter, that decouples when the entire multiplet is heavy, thus a lower bound emerges. The relevant oblique parameters are summarized in~\cite{Lavoura:1993nq} and the global fit was performed in~\cite{Baak:2014ora}.

The resulting limit on $\Delta_L$ is shown on Fig.~\ref{figDeltaLEWPT}, where also the contribution of the $\theta$ mixing to $S$ and $T$ is taken into account. Clearly, the resulting bound on $\Delta_L$ is quite robust and in order to observe $\Delta_L$ at the LHC, a fairly large LR scale is required, beyond the reach of direct $W_R$ searches.

\medskip

\item The charged components of $\Delta_L$ can be pair-produced at the LHC with masses up to about TeV~\cite{Azuelos:2004mwa}. Due to the small $v_L$ (Eq.~\eqref{eqvllimit}) the di-gauge boson final state is suppressed.

When the mass splitting from Eq.~\eqref{eqDLSumRuleLR} is larger than $\mathcal O (\text{GeV})$, the cascade decays $\Delta_L^{++} \to \Delta_L^{+} W^{*+} \to \Delta_L^0 W^{*+} W^{*+}$ open up~\cite{Melfo:2011nx} and necessarily end up in $\Delta_L^0$ due to the sum rule in Eq.~\eqref{eqDLSumRuleLR}. Since $v_L$ is small, c.f. Eq.~\eqref{eqvllimit}, the $W^+ W^-$ and $b \overline b$ final states are suppressed and the resulting final state is $\nu \overline \nu$, i.e. missing energy.

Ultimately, whether $\Delta_L^{++}$ decays via cascades or into same-sign lepton final state depends on the size of the heavy neutrino Majorana Yukawa coupling (see Fig.~1 in~\cite{Melfo:2011nx}).
\end{enumerate}

One might wonder whether a fairly heavy $\Delta_L$ creates an additional perturbativity issue. However, it is clear from Eq.~\eqref{r3} that this is not the case. In fact, for $M_{W_R} \simeq 3 \text{ TeV}$, even $m_{\Delta_L^0} \simeq 13 \text{ TeV}$ is consistent with $\sim 10 \%$ level. Thus the region shown in Fig.~\ref{figDeltaLEWPT} is perturbatively safe as far as $\rho_3$ is concerned.
\vspace*{1em}

%
\subsection{Radiative Higgs decays and $\Delta_R^{++}$} \label{subsecHtogaga}
\noindent
The $\alpha_3$ coupling between the bi-doublet and triplets also induces loop processes with charged particles running in the loop. In particular, this affects the $h \to \gamma \gamma$ rate, whose coupling strength is measured by both ATLAS and CMS~\cite{HiggsCombo}, while $h \to Z \gamma$ is yet to be seen~\cite{Chatrchyan:2013vaa}.

The charged states that couple to the SM Higgs in the LRSM are $W_R, H^+, \Delta_{L}^+$ and $\Delta_{L,R}^{++}$. The RH gauge boson contributions are suppressed due to heavy $W_R$ and $H^+$ is too heavy to contribute, as clear from Table~\ref{scalarP}. As discussed above, the entire $\Delta_L$ multiplet is also heavy, therefore the dominant contribution comes from $\Delta_R^{++}$.

%
There are two contributing amplitudes in radiative channels. One comes from the direct coupling to $h$, the other from interference with $\Delta_R^0$ through the mixing $\theta$.

\begin{figure}
   \includegraphics[width=\columnwidth]{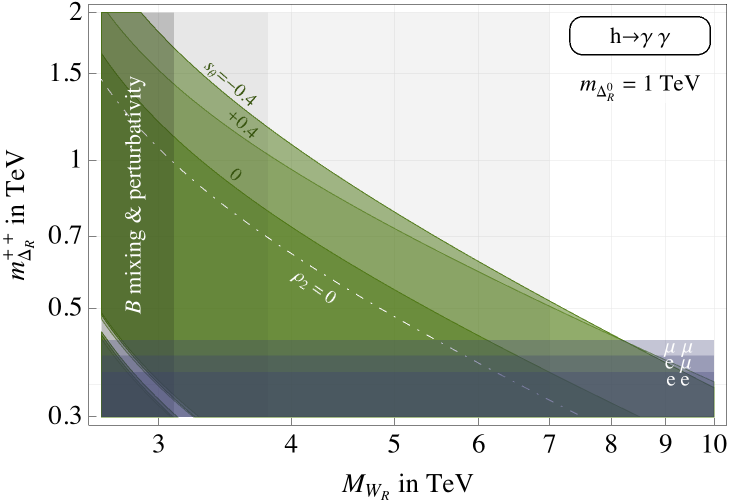}
   \caption{Limit on the mass of $\Delta_R^{++}$ from the current data on $h \to \gamma
     \gamma$~\cite{HiggsCombo}. The shaded green regions are ruled out at $2 \sigma$ level and
     correspond to $s_\theta = (-0.4, 0.4, 0)$ (upper to lower).  The gray vertical bands correspond
     to regions of perturbativity of 10\%, 50\%, 100\% (right to left) as from
     Fig.~\ref{mass_scale}.  The lower blue bands correspond to direct searches for
     $\Delta_R^{++}$~\cite{ATLAS:2014kca} and are flavor dependent. The white strip in the lower
     left part is due to a cancellation with the SM. The region below the white dot-dashed line is
     where $\rho_2$ becomes negative and is disfavored by the boundedness of the classical
     potential.\vspace*{1em}}
  \label{fighgaga}
 \end{figure}

The amplitude is denoted as
\begin{equation}
  \mathcal A^{\gamma \gamma} = \left(\frac{\alpha}{4 \pi}\right) F \,
  \left((p_1 p_2) (\epsilon_1 \epsilon_2) - (\epsilon_1 p_2) (\epsilon_2 p_1) \right) ,
\end{equation}
with $v= 2 \sqrt 2 G_F$. Summing over photon polarizations and including the symmetry factor $1/2$ leads to
\begin{equation}
  \Gamma_{h \to \gamma \gamma} = \left| c_\theta F_h + s_\theta F_\Delta \right|^2 \left( \frac{\alpha}{4 \pi} \right)^2 \frac{m_h^3}{64 \pi}\,.
\end{equation}
The dominant piece coming from $\Delta_R^{++}$ can be extracted from~\eqref{eqAmpDRpp} and one gets
\begin{equation} \label{eqAmpDRpp}
  F_h \simeq F_W + F_f + \sqrt 2 \alpha_3 v Q_{\Delta_R^{++}}^2 F_{\Delta_R^{++}}\,.
\end{equation}
Current limit~\cite{HiggsCombo} on the Higgs $\gamma\gamma$ coupling strength requires the mass of $\Delta_R^{++}$ to be above few 100 GeV, depending on $M_{W_R}$ and the Higgs mixing, as shown in Fig.~\ref{fighgaga}. This constraint turns out to be more relevant than the direct pair-production searches and also overtakes the limit from the boundedness of the potential ($\rho_2 > 0$). It is also relevant for the collider vs. $0\nu2\beta$ connection, since it constrains the rate mediated by $W_R$ and $\Delta_R^{++}$~\cite{Tello:2010am, Nemevsek:2011aa}.

%
The same scalar couplings also enter the $h \to Z \gamma$ loops. With the amplitude defined as
\begin{equation} \label{eqHiggsZga}
  \mathcal A^{Z\gamma}_h = \frac{e g}{\left(4 \pi \right)^2 c_w} G \,
  \left(\left(p_1 p_2\right) \left(\epsilon_1 \epsilon_2\right)  - \left(\epsilon_1 p_2 \right) \left(\epsilon_2 p_1\right) \right),
\end{equation}
the decay rate turns out to be
\begin{equation}
  \!\!\Gamma_{h \to Z \gamma} = \left| c_\theta G_h + s_\theta G_\Delta \right|^2
  \left( \frac{\alpha}{4 \pi} \right) \left( \frac{\alpha_2}{4 \pi} \right) \frac{m_h^3}{32 \pi c_w^2} \beta_{Zh}^3\,,
\end{equation}
where $\beta_{Z h} =  1- M_Z^2/m_h^2$ is the phase space suppression.

Current searches set a limit on the production cross-section around 10x the SM value~\cite{Chatrchyan:2013vaa}. This does not give any additional constraints on charged scalars; in fact even if the coupling strength limit were the same as $\gamma\gamma$, the $Z\gamma$ channel would still be less restrictive.

%
%
\section{Conclusions}

\noindent
The minimal LRSM offers a predictive framework for the origin of neutrino masses and understanding parity violation of the electro-weak interactions. Because of its predictivity, several constraints emerge on the model. In particular, the masses of the FC scalars $H(H')$ have to be large and possibly lead to a tension in the parametric space of the low scale minimal LRSM. This may have a serious impact on the stability and perturbativity of the theory, which we address here.

We systematically study quantum corrections induced by large $\alpha_3$, responsible for the mass of $H$. After reviewing the potential, we perform the one-loop renormalization of the relevant part and build the effective potential.

As a first result, we show that the stability of the potential is improved thanks to the positive loop contribution of $\alpha_3$. This significantly relaxes the stability bound on the mass of the RH neutrino.

A reliable evaluation of the perturbative regime of the LRSM requires a study of the effective potential for all the relevant vertices. By focusing on those, we provide a simple conservative assessment of the regions of parameter space consistent with a given degree of perturbativity.

We translate the above bounds on coupling constants into constraints on physical scalar masses, which we bring together with the known flavor constraints on the LRSM. As a result, we find that a fairly light $W_R$ is compatible with a perturbative effective potential.

Moreover, we study the mixing $\theta$ between the SM Higgs and $\Delta_R^0$, which is a sensitive probe of neutrino mass origin within the LRSM~\cite{Maiezza:2015lza}. We conclude that a significant mixing is perturbatively safe for $\Delta_R^0$ in the TeV range.

We also find that $\alpha_3$ drives the oblique $T$ parameter via $\Delta_L$ mass splitting. Therefore, a light $W_R$ requires a large mass for the entire $\Delta_L$, even in presence of Higgs mixing. We also ensure that this does not introduce an additional perturbativity problem related to the size of the relative quartics $\rho_{3,1}$.

Last but not least, a large $\alpha_3$ modifies the $h \to \gamma \gamma$ decay rate, mainly
through the $\Delta_R^{++}$ loop. In this way, present experimental constraints imply a correlated
lower bound on $m_{\Delta_R^{++}}$ with $M_{W_R}$, which disfavors both of them to be easily
accessible at the LHC, although some borderline space remains.

Notice that as far as the analysis performed in this paper is concerned, the same logic goes through in the model where two doublets~\cite{Senjanovic:1975rk} are considered instead of the usual triplets. Since the quark Yukawa sector is the same, it leads to a large FC Higgs mass that will have a similar issue with perturbativity. It will affect the oblique parameter in a similar way through the splitting of the components of the left-handed doublet and the right-handed singly charged Higgs will enter the $h \to \gamma \gamma$ loop in a similar fashion.

We conclude that a TeV scale $W_R$ requires most of the scalar spectrum to be relatively heavy, apart
from $m_{\Delta_R^0}$, which remains fairly unconstrained even in the presence of mixing.


%
%
\begin{acknowledgments}
\noindent We thank S. Bertolini and G. Senjanovi\'c for useful discussions and reading the manuscript. MN would like to thank B. Bajc for an illuminating discussion. AM was supported in part by the Spanish Government and ERDF funds from the EU Commission  [Grants No. FPA2011-23778, No. CSD2007-00042 (Consolider Project CPAN)] and by Generalitat Valenciana under Grant No. PROMETEOII/2013/007. MN was supported in part by the Slovenian Research Agency. FN was partially supported by the Croatian Science Foundation (HRZZ) project PhySMaB, "Physics of Standard Model and Beyond" as well as by the H2020 Twinning project No. 692194, ``RBI-T-WINNING''.
\end{acknowledgments}

\appendix

\section{Potential(s)} \label{secAppPot}

\vspace*{-1ex}

\noindent
The scalar potential is provided by all the possible bilinear and quartic terms of the scalar
fields, obeying the gauge symmetry and a discrete LR symmetry, $\mathcal{P}$ or $\mathcal{C}$.

Under $\mathcal{P}$, the scalars transform as
\begin{equation}
  \mathcal P: \phi \to \phi^\dagger, \, \Delta_L \leftrightarrow \Delta_R
\end{equation}
and for the potential one has
\begin{widetext}
\begin{align}
\label{eqVP}
\mathcal{V_P} &= -
\mu_1^2 \left[ \phi ^\dagger \phi \right] -
\mu_2^2 \left( \left[\tilde{\phi} \phi ^\dagger \right] + \left[ \tilde{\phi }^\dagger \phi \right]\right) -
\mu _3^2 \left( \left[\Delta_L^{{\color{white}\dagger}} \Delta _L^\dagger\right] + \left[\Delta_R^{{\color{white}\dagger}} \Delta_R^\dagger\right]\right)
\nonumber\\[-.5ex]
&+ \lambda_1 \left[\phi ^\dagger \phi \right]^2 +
\lambda_2 \left(\left[\tilde{\phi } \phi ^\dagger\right]^2 + \left[\tilde{\phi }^\dagger \phi \right] ^2\right) +
\lambda_3 \left[\tilde{\phi } \phi ^\dagger\right] \left[\tilde{\phi }^\dagger \phi \right]
+ \lambda_4 \left[\phi ^\dagger \phi \right] \left(\left[\tilde{\phi } \phi ^\dagger\right] + \left[\tilde{\phi }^\dagger \phi \right]\right)
\nonumber\\[-.5ex]
&+ \rho_1 \left(\left[\Delta_L^{{\color{white}\dagger}} \Delta _L^\dagger\right]^2 + \left[\Delta_R^{{\color{white}\dagger}} \Delta_R^\dagger\right]^2\right)
+ \rho_2 \left(\left[\Delta_L^{{\color{white}\dagger}} \Delta _L^{{\color{white}\dagger}} \right]
\left[\Delta _L^\dagger \Delta _L^\dagger\right]+\left[\Delta_R^{{\color{white}\dagger}} \Delta_R^{{\color{white}\dagger}} \right] \left[\Delta _R^\dagger \Delta _R^\dagger\right]\right)
+ \rho_3 \left[\Delta_L^{{\color{white}\dagger}} \Delta _L^\dagger\right] \left[\Delta_R^{{\color{white}\dagger}} \Delta _R^\dagger\right]
\nonumber\\[-.5ex]
&+ \rho _4 \left(\left[\Delta_L^{{\color{white}\dagger}} \Delta _L^{{\color{white}\dagger}} \right] \left[\Delta_R^\dagger \Delta _R^\dagger\right] + \left[\Delta _L^\dagger \Delta _L^\dagger\right] \left[\Delta_R^{{\color{white}\dagger}} \Delta_R^{{\color{white}\dagger}} \right]\right)
 + \alpha _1 \left[\phi ^\dagger \phi \right] \left(\left[\Delta_L^{{\color{white}\dagger}} \Delta _L^\dagger\right]+\left[\Delta_R^{{\color{white}\dagger}} \Delta _R^\dagger\right]\right)
\\
&+ \alpha _2 e^{i \delta _2} \left(\left[\tilde{\phi } \phi ^\dagger\right] \left[\Delta_L^{{\color{white}\dagger}} \Delta _L^\dagger\right]+\left[\tilde{\phi }^\dagger \phi \right] \left[\Delta_R^{{\color{white}\dagger}} \Delta _R^\dagger\right]\right) + \text{h.c.}
\nonumber\\
&
+\alpha _3 \left(\left[\phi  \phi ^\dagger \Delta_L^{{\color{white}\dagger}} \Delta _L^\dagger\right]+\left[\phi ^\dagger \phi  \Delta_R^{{\color{white}\dagger}} \Delta _R^\dagger\right]\right)
\nonumber\\
&
 + \beta _1 \left(\left[\phi  \Delta_R^{{\color{white}\dagger}} \phi ^\dagger \Delta _L^\dagger\right]+\left[\phi ^\dagger \Delta_L^{{\color{white}\dagger}} \phi  \Delta _R^\dagger\right]\right)+\beta _2 \left(\left[\tilde{\phi } \Delta_R^{{\color{white}\dagger}} \phi ^\dagger \Delta_L^\dagger\right]+\left[\tilde{\phi }^\dagger \Delta_L^{{\color{white}\dagger}} \phi  \Delta _R^\dagger\right]\right)
+ \beta _3 \left(\left[\phi  \Delta_R^{{\color{white}\dagger}} \tilde{\phi}^\dagger \Delta _L^\dagger\right]+\left[\phi ^\dagger \Delta_L^{{\color{white}\dagger}} \tilde{\phi } \Delta _R^\dagger\right]\right),
\nonumber
\end{align}
\end{widetext}
where $\tilde{\phi}= \sigma_2 \phi^{*}  \sigma_2$ and the square brackets imply the trace over field components.

From the minimization conditions in Eq.~\eqref{eqDV}, the $\mu_i$ are computed in terms of vevs and the quartics
\begin{align}\label{eqmu12P}
\begin{split}
  \mu_1^2 =&\; \frac{2 v_1^2 \lambda_1 \left(x^4-1\right)+4 v_1^2 \lambda_4 x \left(x^2-1\right) \cos (\alpha )}{x^2-1}
  \\
  &{}+ \frac{v_R^2 \left(\alpha_1 \left(x^2-1\right)+\alpha_3 x^2\right)}{x^2-1}\,,
\end{split}
\\
\label{eqmu22P}
\mu_2^2 =&\; \sec (\alpha )\,\bigg[
2 v_1^2 \lambda_4 \left(x^2+1\right) \cos (\alpha )
-\frac{\alpha_3 v_R^2 x}{2\left(x^2-1\right)}
\\
&{}+  4 v_1^2 \lambda_2 x \cos (2 \alpha )
+2 v_1^2 \lambda_3 x + \alpha_2 v_R^2 \cos (\alpha +\delta_2)\bigg]\,
\nonumber
\\
\label{eqmu32P}
\begin{split}
  \mu_3^2 =&\; 2 \rho_1 v_R^2 + \left(\alpha_1+ (\alpha_1+\alpha_3) x^2\right) v_1^2
  \\
  &{}+ 4 \alpha_2  x \cos (\alpha +\delta_2) v_1^2\,,
\end{split}
\end{align}
where $v_L$ is neglected, since $v_L \ll v \ll v_R$.

 The derivative over $\alpha$ requires the relation between CP phases:
\begin{align} \label{eqRelD2}
  \sin(\delta_2)&= \frac{x \sin (\alpha)}{2 \alpha_2 }
  \left[\frac{\alpha_3}{1- x^2} - 4 \epsilon ^2 \frac{2 \lambda_2-\lambda_3}{1+x^2}\right].
\end{align}

\medskip

Alternatively, one can adopt the $\mathcal{C}$ symmetry, which acts as
\begin{equation}
  \mathcal C : \phi \to \phi^T\,, \qquad \Delta_L \leftrightarrow \Delta_R^*\,.
\end{equation}
This choice allows for additional complex phases in the potential~\cite{Dekens:2014ina}.  In
particular, the couplings $\mu_2$, $\lambda_2$, $\lambda_4$, $\rho_4$ and $\beta_i$ are now
complex and one has to introduce a phase $\delta_c$ for each such coupling $c$. The same applies to
the Yukawa sector leading to additional free phases in the RH CKM matrix~\cite{Maiezza:2010ic, Senjanovic:2014pva}.

Again, the minimization conditions provide the $\mu_i$ parameters
\begin{align}
\begin{split}
\mu_1^2 &= \frac{1}{x^2-1} \biggl[
4 v_1^2 \lambda_4 x \left(x^2-1\right) \cos (\alpha -\delta_{\lambda_{4}})
\\
&+ 2 v_1^2 \lambda_1 \left(x^4-1\right) + v_R^2 \left(\alpha_1 \left(x^2-1\right)+\alpha_3 x^2\right) \biggr]\,,
\end{split}
\\
\begin{split}
\mu_2^2 &= \frac{\sec (\alpha -\delta_{\mu_{2}})}{2 \left(x^2-1\right)} \biggl[
(x^4 - 1) 2 v_1^2 \lambda_4 \cos (\alpha -\delta_{\lambda_{4}})
\\
&+ 4 v_1^2 x(x^2-1) (2 \lambda_2 \cos (2 \alpha -\delta_{\lambda_{2}}) + \lambda_3)
\\
&+ 2 \alpha_2 v_R^2 x^2 \cos (\alpha -\delta_2)(x^2-1) -\alpha_3 v_R^2 x \biggr]\,,
\end{split}
\\
\begin{split}
  \mu_3^2 &= 2 \rho_1 v_R^2 + v_1^2 \left(\alpha_1 (1+ x^2) + \alpha_3 x^2\right)
  \\
  &+4 \alpha_2 v_1^2 x \cos (\alpha -\delta_2)\,,
\end{split}
\end{align}
together with Equations analogous to~\eqref{eqVevssaw} and \eqref{eqRelD2}:
\begin{align}
 &v_L = \varepsilon^2 v_R \frac{\beta_1 x \cos (\alpha +\delta_ {\beta_{1}}-\theta_L)
 }{\left(1 + x^2\right) (2 \rho_1-\rho_3)}
 \\
 &\qquad+ \frac{\beta_2 \cos (\delta_{\beta_{2}}-\theta_L) + \beta_3 x^2 \cos (2 \alpha +\delta_{\beta_{3}} - \theta_L)}{\left(1 + x^2\right) (2 \rho_1-\rho_3)}\,,
\nonumber
  \\
  &\sin(\delta_2-\delta_{\mu_{2}}) \simeq  \frac{\alpha_3 x \sin (\alpha -\delta_{\mu_{2}})}{2 \alpha_2 \left(x^2-1\right)}\,.
\end{align}
The masses of the FC scalars in the case of $\mathcal{C}$ are
\begin{equation}
  m_{H(H')}^2 = v_R^2 \left[\alpha_3 +8\,\varepsilon^2\left(\frac{ \alpha_2^2 \pm  \tilde \alpha}{\alpha_3-4 \rho_1}+ \frac{\lambda_3}{2}\right)\right],
\end{equation}
where
\begin{equation}
\begin{split}
  \tilde \alpha^2 =&\; \alpha_2^4 + \lambda_2 (\alpha_3-4 \rho_1) \bigl[  \lambda_2 (\alpha_3-4 \rho_1) +
  \\
  & + 2 \alpha_2^2 \cos (2 \delta_2-\delta_{\lambda _{2}}) \bigr],
\end{split}
\end{equation}
and which reproduces the case of $\mathcal{P}$ for vanishing extra-phases. A Left-Right potential with no $L\leftrightarrow R$ discrete symmetry was discussed in~\cite{Dekens:2014ina}.

%
%
\section{Higgs decay loop functions} \label{AsecHiggsRadDec}

\paragraph{Higgs to $\gamma \gamma$.} The SM contribution to the di-photon Higgs decay was computed some time ago~\cite{HiggsToGaGa}. The loop coefficients are
\begin{align}
  F_h &= \sum_f N_f Q_f^2 F_f + F_W + \sum_S c^h_S v \, Q_S^2 F_S\,,
  \\
  F_\Delta &= F_{W_R} + \sum_S c^\Delta_S v_R \, Q_S^2 F_S\,,
\end{align}
and
\begin{align}
  F_f    &= -\frac{\sqrt{2}}{v} 2 \beta_f \left[1 + \left(1 - \beta_f \right) f(\beta_f)\right],
  \\
  F_W &= \frac{\sqrt{2}}{v} \left[2 + 3 \beta_W \left(1 + (2 - \beta_W) f(\beta_W)) \right) \right],
  \\
  F_S   &= \frac{\beta_S}{m_S^2} \left[1 - \beta_S f(\beta_S) \right],
  \\
  F_{W_R} &= \frac{v}{v_R} F_W(\beta_{W_R})\,.
\end{align}
The corresponding dimensionless couplings $c_S^{h,\Delta}$ are obtained from the potential and are listed in Table~\ref{tabhDcpl}. The dimensionless $f$ is the usual
\begin{align}
  f (\beta \geq 1) &= \arcsin \left(1/\sqrt \beta \right)^2,
  \\
  f (\beta < 1) &= -\frac{1}{4} \left( \log \left( \frac{1 + \sqrt{1 - \beta}}{1 - \sqrt{1 - \beta}} \right) - i \pi \right)^2,
\end{align}
and $\beta_i = (2 M_i/m_h)^2$. The scalar exchange part is
\begin{equation}
\label{eqAmpDRpp}
\begin{split}
  F_h + F_\Delta &=
  \left( c_\theta c^h_S v + s_\theta c^{\Delta}_S v_R \right) Q_S^2 F_S
  \\
  & \xrightarrow{\alpha_1 \to 0} \sqrt 2 \alpha_3 v Q_S^2 F_S\,.
\end{split}
\end{equation}
This is typically the dominant new physics piece, since $\alpha_1$ should not be too large for the theory to remain perturbative. For the same reason, we expect the second term in Eq.~\eqref{eqAmpDRpp} to be subdominant, also due to the mixing angle suppression.

\begin{table}[t] \centering
\renewcommand{\arraystretch}{1.5}
$\begin{array}{| c | c c c |}
  \hline
		& \Delta_R^{++} & \Delta_L^{++} (\Delta_L^+) & H^+
  \\ \hline
  c^h_S	& \sqrt 2 (\alpha_1 + \alpha_3) & \sqrt 2 (\alpha_1 + \alpha_3)  & 2 \sqrt 2 \lambda_1
  \\
  c^\Delta_S 	& 2 \sqrt 2 (\rho_1 + 2 \rho_2) & \sqrt 2 \rho_3 & \sqrt 2 (\alpha_1 + \alpha_3)
  \\ \hline	
\end{array}$
\caption{Couplings of the LRSM Higgs bosons to charged scalars, relevant for radiative decays. $S$ denotes the charged scalars  $S = (\Delta_R^{++}, \Delta_L^{++}, \Delta_L^+, H^+)$.}
\label{tabhDcpl}
\end{table}

\paragraph{Higgs to $Z \gamma$.} The amplitude coefficients are
\begin{align}
\begin{split}
  G_h =&\; \sum_f N_f Q_f \hat v_f G_f + G_W
  \\
  & + \sum_S c^h_S v \, Q_S \hat v_S \, G_S\,,
\end{split}
  \\
  G_\Delta =&\; G_{W_R} + \sum_S  c^{\Delta}_S v_R  \, Q_S \hat v_S \, G_S\,,
\end{align}
where $\hat v_f = T_{3L}/2 - Q s_w^2$. The individual pieces are
\begin{equation}
  \begin{split}
  G_f  =&\; \frac{2 \sqrt 2 \beta \lambda}{v (\beta - \lambda)^2}  \biggl[\beta - \lambda + ((\beta - 1) \lambda + \beta)
  \\
  &\times(f(\beta) - f(\lambda)) - 2 \beta (g(\lambda) - g(\beta)) \biggr]\,,
  \end{split}
\end{equation}
\phantom{Don't look this Don't look this Don't look this}

\begin{equation}
  \begin{split}
  G_W =&\; \frac{\sqrt 2 c_w^2}{v (\beta - \lambda)^2} \biggl[
  (4 + 2 (\beta - \lambda) - 3 \beta \lambda) (\beta (1 +
  \\
  &2 (g(\beta) - g(\lambda))) - \lambda) - \beta \left(f(\beta) - f(\lambda) \right)
  \\
  & \times\left(\beta (3 \lambda (\lambda + 2) - 8) + 2 (2 - 3 \lambda) \lambda \right) \biggr]\,,
  \end{split}
\end{equation}

\begin{equation}
\begin{split}
  G_S   =&\; \frac{\beta \lambda}{m_S^2 \left( \beta - \lambda \right)^2} \biggl[ \lambda - \beta + \beta (\lambda \left( f(\lambda) - f(\beta) \right)
  \\
  & + 2 ( g(\lambda) - g(\beta)))\biggr]\,,
\end{split}
\end{equation}

\begin{equation}
  G_{W_R} = \frac{v}{v_R} G_W(\beta_{W_R})\,.
\end{equation}
where
\begin{align}
  g (\beta \leq 1) &= \sqrt{\beta - 1} \arcsin \left(1/\sqrt \beta \right),
  \\
  g (\beta > 1) &= \frac{\sqrt{1-\beta}}{2} \left( \log \left( \frac{1 + \sqrt{1 - \beta}}{1 - \sqrt{1 - \beta}} \right) - i \pi \right).
\end{align}
In the $M_Z \to 0$ limit the $F$ functions are recovered
\begin{equation}
  G_i \xrightarrow{\lambda_i \to \infty} F_i\,.
\end{equation}

\medskip

\def\arxiv#1[#2]{\href{http://arxiv.org/abs/#1}{[#2]}}
\def\Arxiv#1[#2]{\href{http://arxiv.org/abs/#1}{#2}}


\begin{thebibliography}{99}

\bibitem{Pati:1974yy}
  J.C.~Pati and A.~Salam,
  Phys.\ Rev.\ D {\bf 10}, 275 (1974)
  [Erratum-ibid.\ D {\bf 11}, 703 (1975)];
  R.N.~Mohapatra and J.C.~Pati,
  Phys.\ Rev.\ D {\bf 11}, 566 (1975);
  R.N.~Mohapatra and J.C.~Pati,
  Phys.\ Rev.\ D {\bf 11}, 2558 (1975).

\bibitem{Senjanovic:1975rk}
  G.~Senjanovi\'c and R.N.~Mohapatra,
  Phys.\ Rev.\ D {\bf 12}, 1502 (1975);
  G.~Senjanovi\'c,
  Nucl.\ Phys.\ B {\bf 153}, 334 (1979).

\bibitem{MohSen}
R.N.~Mohapatra, G.~Senjanovi\'{c},
Phys.\ Rev.\ Lett. {\bf 44} (1980) 912.

\bibitem{Minkowski}
P.~Minkowski,
Phys.\ Lett.\ B {\bf 67} (1977) 421.

\bibitem{Mohapatra:1980yp}
  R.N.~Mohapatra and G.~Senjanovi\'c,
  Phys.\ Rev.\ D {\bf 23} (1981) 165.

\bibitem{Tello:2010am}
  V.~Tello, M.~Nemev\v{s}ek, F.~Nesti, G.~Senjanovi\'c and F.~Vissani,
  Phys.\ Rev.\ Lett.\  {\bf 106} (2011) 151801
  \arxiv{1011.3522}[arXiv:1011.3522 [hep-ph]].

\bibitem{Nemevsek:2011aa}
  M.~Nemev\v{s}ek, F.~Nesti, G.~Senjanovi\'c and V.~Tello,
  \arxiv{1112.3061}[arXiv:1112.3061 [hep-ph]].

\bibitem{Keung:1983uu}
  W.-Y.~Keung, G.~Senjanovi\'c,
  Phys.\ Rev.\ Lett.\  {\bf 50 } (1983)  1427.

\bibitem{GSreview} For a review, see:
  G.~Senjanovi\'c,
  Int.\ J.\ Mod.\ Phys.\ A {\bf 26} (2011) 1469
  [arXiv:1012.4104 [hep-ph]];
  G.~Senjanovi\'c,
  Riv.\ Nuovo Cim.\  {\bf 34} (2011) 1.


\bibitem{Gunion:1986im}
 J.F.~Gunion, B.~Kayser, R.N.~Mohapatra, N.G.~Deshpande, J.~Grifols, A.~Mendez, F.I.~Olness and P.B.~Pal,
  PRINT-86-1324 (UC,DAVIS);
  J.~F.~Gunion, H.~E.~Haber, G.~L.~Kane and S.~Dawson,
  Front.\ Phys.\  {\bf 80} (2000) 1.

\bibitem{Maiezza:2015lza}
  A.~Maiezza, M.~Nemev\v{s}ek and F.~Nesti,
  Phys.\ Rev.\ Lett.\  {\bf 115} (2015) 081802
  \arxiv{1503.06834}[arXiv:1503.06834 [hep-ph]].

\bibitem{Maiezza:2010ic}
  A.~Maiezza, M.~Nemev\v{s}ek, F.~Nesti and G.~Senjanovi\'c,
  Phys.\ Rev.\ D {\bf 82}, 055022 (2010)
  \arxiv{1005.5160}[arXiv:1005.5160 [hep-ph]].

\bibitem{Nemevsek:2012iq}
  M.~Nemev\v{s}ek, G.~Senjanovi\'c and V.~Tello,
  Phys.\ Rev.\ Lett.\  {\bf 110} (2013) 15,  151802
  \arxiv{1211.2837}[arXiv:1211.2837 [hep-ph]].

\bibitem{Maiezza:2014ala}
  A.~Maiezza and M.~Nemev\v{s}ek,
  Phys.\ Rev.\ D {\bf 90} (2014) 9,  095002
  \arxiv{1407.3678}[arXiv:1407.3678 [hep-ph]].

\bibitem{Senjanovic:2014pva}
  G.~Senjanovi\'c and V.~Tello,
  Phys.\ Rev.\ Lett.\  {\bf 114} (2015) 7,  071801
  \arxiv{1408.3835}[arXiv:1408.3835 [hep-ph]],
  \arxiv{1502.05704}[arXiv:1502.05704 [hep-ph]].

\bibitem{Beall:1981ze}
  G.~Beall, M.~Bander and A.~Soni,
  Phys.\ Rev.\ Lett.\  {\bf 48} (1982) 848.

\bibitem{Ecker:1985vv}
  G.~Ecker and W.~Grimus,
  Nucl.\ Phys.\  B {\bf 258}, 328 (1985).

\bibitem{Zhang:2007da}
  Y.~Zhang, H.~An, X.~Ji and R.N.~Mohapatra,
  Nucl.\ Phys.\ B {\bf 802}, 247 (2008)
  \arxiv{0712.4218}[arXiv:0712.4218 [hep-ph]].

\bibitem{Bertolini:2014sua}
  S.~Bertolini, A.~Maiezza and F.~Nesti,
  Phys.\ Rev.\ D {\bf 89} (2014) 9, 095028
  \arxiv{1403.7112}[arXiv:1403.7112 [hep-ph]].

\bibitem{Bertolini:2012pu}
  S.~Bertolini, J.O.~Eeg, A.~Maiezza and F.~Nesti,
  Phys.\ Rev.\ D {\bf 86} (2012) 095013
  \arxiv{1206.0668}[arXiv:1206.0668 [hep-ph]];
  S.~Bertolini, A.~Maiezza and F.~Nesti,
  Phys.\ Rev.\ D {\bf 88} (2013) 3,  034014
  \arxiv{1305.5739}[arXiv:1305.5739 [hep-ph]].

\bibitem{Senjanovic:1979cta}
  G.~Senjanovi\'c and P.~Senjanovi\'c,
  Phys.\ Rev.\ D {\bf 21}, 3253 (1980).

\bibitem{Guadagnoli:2010sd}
  D.~Guadagnoli and R.~N.~Mohapatra,
  Phys.\ Lett.\ B {\bf 694}, 386 (2011)
  \arxiv{arXiv:1008.1074}[arXiv:1008.1074 [hep-ph]].

\bibitem{Gunion:1989in}
  J.~F.~Gunion, J.~Grifols, A.~Mendez, B.~Kayser and F.~I.~Olness,
  Phys.\ Rev.\ D {\bf 40} (1989) 1546.

\bibitem{Kiers:2002cz}
  K.~Kiers, J.~Kolb, J.~Lee, A.~Soni and G.~H.~Wu,
  Phys.\ Rev.\ D {\bf 66}, 095002 (2002)
  \arxiv{hep-ph/0205082}[hep-ph/0205082].

\bibitem{Deshpande:1990ip}
  N.~G.~Deshpande, J.~F.~Gunion, B.~Kayser and F.~I.~Olness,
  Phys.\ Rev.\ D {\bf 44} (1991) 837.

\bibitem{Duka:1999uc}
  P.~Duka, J.~Gluza and M.~Zra\l ek,
  Annals Phys.\  {\bf 280} (2000) 336
  \arxiv{hep-ph/9910279}[hep-ph/9910279].

\bibitem{Khasanov:2001tu}
  O.~Khasanov and G.~Perez,
  Phys.\ Rev.\ D {\bf 65} (2002) 053007
  \arxiv{hep-ph/0108176}[hep-ph/0108176];
  K.~Kiers, M.~Assis and A.~A.~Petrov,
  Phys.\ Rev.\ D {\bf 71} (2005) 115015
  \arxiv{hep-ph/0503115}[hep-ph/0503115];
  K.~Kiers, M.~Assis, D.~Simons, A.~A.~Petrov and A.~Soni,
  Phys.\ Rev.\ D {\bf 73} (2006) 033009
  \arxiv{hep-ph/0510274}[hep-ph/0510274].

\bibitem{Dekens:2014ina}
  W.~Dekens and D.~Boer,
  Nucl.\ Phys.\ B {\bf 889}, 727 (2014)
  \arxiv{1409.4052}[arXiv:1409.4052 [hep-ph]].

\bibitem{Bambhaniya:2015wna}
  G.~Bambhaniya, J.~Chakrabortty, J.~Gluza, T.~Jelinski and R.~Szafron,
  Phys.\ Rev.\ D {\bf 92} (2015) 1,  015016
  \arxiv{1504.03999}[arXiv:1504.03999 [hep-ph]].

\bibitem{Dev:2016dja}
  P.~S.~B.~Dev, R.~N.~Mohapatra and Y.~Zhang,
  \arxiv{1602.05947}[arXiv:1602.05947 [hep-ph]].


\bibitem{Barbieri:1988av}
  R.~Barbieri and R.~N.~Mohapatra,
  Phys.\ Rev.\ D {\bf 39} (1989) 1229.

\bibitem{Nemevsek:2012cd}
  M.~Nemev\v{s}ek, G.~Senjanovi\'c and Y.~Zhang,
  JCAP {\bf 1207} (2012) 006
  \arxiv{1205.0844}[arXiv:1205.0844 [hep-ph]].

\bibitem{GoranJuan16}
  G.~Senjanovi\'c, J.~C.~Vasquez,
  work in progress.

\bibitem{Basecq:1985cr}
  J.~Basecq, L.~F.~Li and P.~B.~Pal,
  Phys.\ Rev.\ D {\bf 32}, 175 (1985).

\bibitem{Peskin} M.E.~Peskin and D.V.~Schroeder,
  An Introduction To Quantum Field Theory (Frontiers in Physics).

\bibitem{Mohapatra:1986pj}
  R.~N.~Mohapatra,
  Phys.\ Rev.\ D {\bf 34} (1986) 909.


\bibitem{Chakrabortty:2013mha}
  J.~Chakrabortty, P.~Konar and T.~Mondal,
  Phys.\ Rev.\ D {\bf 89} (2014) 5,  056014
  \arxiv{1308.1291}[arXiv:1308.1291 [hep-ph]].
  Phys.\ Rev.\ D {\bf 89} (2014) 9,  095008
  \arxiv{1311.5666}[arXiv:1311.5666 [hep-ph]].

\bibitem{UseOfCopositivity}
  G.~Bambhaniya, J.~Chakrabortty, J.~Gluza, T.~Jelinski and R.~Szafron,
  Phys.\ Rev.\ D {\bf 92} (2015) 1,  015016
  \arxiv{1504.03999}[arXiv:1504.03999 [hep-ph]];
  A.~Karam and K.~Tamvakis,
  Phys.\ Rev.\ D {\bf 92} (2015) 7,  075010
  \arxiv{1508.03031}[arXiv:1508.03031 [hep-ph]];
  T.~Mondal, U.~K.~Dey and P.~Konar,
  Phys.\ Rev.\ D {\bf 92} (2015) 9,  096005
  \arxiv{1508.04960}[arXiv:1508.04960 [hep-ph]];
  N.~Haba and Y.~Yamaguchi,
  PTEP {\bf 2015} (2015) 9,  093B05
  \arxiv{1504.05669}[arXiv:1504.05669 [hep-ph]].

\bibitem{singletHiggs}
  A.~Falkowski, C.~Gross and O.~Lebedev,
  JHEP {\bf 1505} (2015) 057
  \arxiv{1502.01361}[arXiv:1502.01361 [hep-ph]];
  S.~I.~Godunov, A.~N.~Rozanov, M.~I.~Vysotsky and E.~V.~Zhemchugov,
  Eur.\ Phys.\ J.\ C {\bf 76} (2016) 1,  1
  doi:10.1140/epjc/s10052-015-3826-6
  \arxiv{1503.01618}[arXiv:1503.01618 [hep-ph]].

\bibitem{Melfo:2011nx}
  A.~Melfo, M.~Nemev\v{s}ek, F.~Nesti, G.~Senjanovi\'c and Y.~Zhang,
  Phys.\ Rev.\ D {\bf 85} (2012) 055018
  \arxiv{1108.4416}[arXiv:1108.4416 [hep-ph]].

\bibitem{Lavoura:1993nq}
  L.~Lavoura and L.~F.~Li,
  Phys.\ Rev.\ D {\bf 49} (1994) 1409
  \arxiv{9309262}[hep-ph/9309262].

\bibitem{Baak:2014ora}
  M.~Baak {\it et al.} [Gfitter Group Collaboration],
  Eur.\ Phys.\ J.\ C {\bf 74} (2014) 3046
  \arxiv{1407.3792}[arXiv:1407.3792 [hep-ph]].

\bibitem{Azuelos:2004mwa}
  G.~Azuelos, K.~Benslama and J.~Ferland,
  J.\ Phys.\ G {\bf 32} (2006) 2,  73
  \arxiv{0503096}[hep-ph/0503096].

\bibitem{HiggsCombo}
  The ATLAS and CMS Collaborations,
  ATLAS-CONF-2015-044.

\bibitem{Chatrchyan:2013vaa}
  S.~Chatrchyan {\it et al.} [CMS Collaboration],
  Phys.\ Lett.\ B {\bf 726} (2013) 587
  \arxiv{1307.5515}[arXiv:1307.5515 [hep-ex]].

\bibitem{ATLAS:2014kca}
  G.~Aad {\it et al.} [ATLAS Collaboration],
  JHEP {\bf 1503} (2015) 041
  \arxiv{1412.0237}[arXiv:1412.0237 [hep-ex]].



\bibitem{HiggsToGaGa}
    J.~R.~Ellis, M.~K.~Gaillard and D.~V.~Nanopoulos,
  Nucl.\ Phys.\ B {\bf 106} (1976) 292;
  M.~A.~Shifman, A.~I.~Vainshtein, M.~B.~Voloshin and V.~I.~Zakharov,
  Sov.\ J.\ Nucl.\ Phys.\  {\bf 30} (1979) 711
   [Yad.\ Fiz.\  {\bf 30} (1979) 1368];
  W.~J.~Marciano, C.~Zhang and S.~Willenbrock,
  Phys.\ Rev.\ D {\bf 85} (2012) 013002
  \arxiv{1109.5304}[arXiv:1109.5304 [hep-ph]].

\end{thebibliography}
\end{document}